\documentclass[conference]{IEEEtran}
\usepackage{url}

\ifCLASSINFOpdf

\else

\fi

\begin{document}

\title{Mull it over: mutation testing based on LLVM}

\author{\IEEEauthorblockN{Alex Denisov}
\IEEEauthorblockA{Independent researcher\\
Berlin, Germany\\
Email: alex@lowlevelbits.org}
\and
\IEEEauthorblockN{Stanislav Pankevich}
\IEEEauthorblockA{Independent researcher\\
Berlin, Germany\\
Email: s.pankevich@gmail.com}}

\maketitle

\begin{abstract}
This paper describes Mull, an open-source tool for mutation testing based on the LLVM framework. Mull works with LLVM IR, a low-level intermediate representation, to perform mutations, and uses LLVM JIT for just-in-time compilation. This design choice enables the following two capabilities of Mull: language independence and fine-grained control over compilation and execution of a tested program and its mutations. Mull can work with code written in any programming language that supports compilation to LLVM IR, such as C, C++, Rust, or Swift. Direct manipulation of LLVM IR allows Mull to do less work to generate mutations: only modified fragments of IR code are recompiled, and this results in faster processing of mutated programs. To our knowledge, no existing mutation testing tool provides these capabilities for compiled programming languages. We describe the algorithm and implementation details of Mull, highlight current limitations of Mull, and present the results of our evaluation of Mull on real-world projects such as RODOS, OpenSSL, LLVM.
\end{abstract}


\IEEEpeerreviewmaketitle

\section{Introduction}

Mutation Testing, a fault-based software testing technique, serves as a way to evaluate
and improve quality of software tests. A tool for mutation testing creates many slightly
modified versions of original program and then runs a test suite
against each version, which is called a \emph{mutant}. A mutant is said to be \emph{killed} if the test suite
detects a change to the program introduced by this mutant,
or \emph{survived} otherwise. Each mutation of original program is created based on 
one of the predefined rules for program modification called \emph{mutation operators}. 
Each mutant is represented by a \emph{mutation point}: a combination of mutation operator 
and location of a mutation in the program's source code. To assess the quality of a test suite 
mutation testing uses a metric called \emph{mutation score}, or \emph{mutation coverage}.

Mutation testing is getting interest from the open source community. More and more open-source mutation testing tools targeting various programming languages appear \cite{GithubTopicsMutationTesting}. Unfortunately, not all of these tools reach a level of maturity needed for practical use. While mature implementations of open-source mutation testing tools definitely exist, with Pitest \cite{PITEST} and Mutant \cite{MUTANT} being strong examples from Java and Ruby programming language communities, there is still a lack of usable mutation testing tools for certain compiled programming languages.

In this paper, we present the Mull project, our attempt to build a general-purpose mutation testing tool targeting compiled languages. Mull is built on top of the LLVM compiler framework \cite{LLVMPaper}. It uses two components of LLVM: IR, its low-level intermediate language, to perform mutations and JIT for runtime compilation and execution of a tested program and its mutated counterparts. LLVM IR is also referred to as \emph{LLVM Bitcode} or simply as \emph{bitcode}. We use these terms interchangeably.

We consider the following criteria important for a practical implementation of mutation testing tool: the tool must be fast, configurable and easy to set up and use. The tool should allow smooth integration with build tools. The tool should be ready for use in mutation testing analysis of real-world production and open source projects. The tool should implement a reasonable number of basic mutation operators to enable the practical use of it in different domains such as systems programming, application programming, algorithms and mathematical computations.

Mull is built with all of the above criteria in mind. We started Mull with a primary focus
on C and C++, but due to LLVM, Mull can work with any other programming language that
compiles to LLVM IR, such as Rust, Swift, Objective-C. To add a language support one needs
to implement adapters to the test frameworks used by the programming language.

Mull is a command line tool. It takes a configuration file as an input and produces
an SQLite database with the results as output. Configuration options include a list of tested 
program's bitcode files, a set of mutation operators, a test framework, and a few other
settings. The SQLite database contains information that Mull gathers while running on a
tested program, such as tests, mutation points, mutants (killed or survived), and more.
As a command-line tool, Mull does not show mutation score or mutation coverage. There is a separate program that generates an HTML report from the SQLite file.

Mull's source code is available online \cite{MullSourceCode} under Apache License, version 2.0 \cite{apache2}.

We organize the rest of the paper as follows. Section II describes the algorithm of Mull. Section III then goes deeper and describes what we consider the most interesting implementation details of Mull. Section IV describes the mutation operators currently implemented in Mull. Section V describes our evaluation of the open source projects: RODOS, OpenSSL, LLVM. Section VI highlights the limitations of Mull. Section VII discusses future work. Section VIII concludes the paper.

\section{Algorithm}

The following are the steps that Mull performs during a session:

\subsubsection*{Step 1} Mull loads LLVM Bitcode into memory.

\subsubsection*{Step 2} Mull inserts instrumentation code into each function. This code is used to collect code coverage information. We describe our approach to instrumentation in III.A.

\subsubsection*{Step 3} Mull compiles instrumented LLVM Bitcode to machine code and prepares the machine code for execution by LLVM JIT engine.

\subsubsection*{Step 4} In the LLVM IR code Mull finds the tests according to a test framework specified in the configuration file.

\subsubsection*{Step 5} Mull runs each test using LLVM JIT engine and collects code coverage information.

\subsubsection*{Step 6} Mull finds mutations in the LLVM IR code based on a code coverage information collected for each test. A set of mutation points is created.

\subsubsection*{Step 7} For each mutation point, Mull creates a mutant and runs each test that can kill the mutant. For each mutant, only part of bitcode is recompiled into machine code. We describe our approach to runtime compilation in III.B.

\subsubsection*{Step 8} All information collected during the session is written to the
SQLite database. This is the final step. Mull finishes its execution at this point.

\section{Implementation}

\subsection{Instrumentation and Dynamic Call Tree} \label{dynamic_call_tree}

A typical program has many mutations, but not all of them are reachable by the program's tests.
We use this fact to reduce the number of mutants. To know which mutations are reachable we thus need to know which code is reachable from a test. To achieve this, we insert instrumentation into each function and then run a test to gather code coverage information. From this information, we construct a \emph{dynamic call tree}.

The purpose of the call tree is better illustrated by example. Consider a function \verb;test_driver; calling function \verb;test; which is calling functions \verb;testee1; and \verb;testee2;. The call tree would look like the following: \verb;test_driver -> test -> { testee1, testee2 };. In this case the code being tested is inside of \verb;testee1; and \verb;testee2;. Therefore we can inject mutations only into the subtrees of the \verb;test; function.

The call tree adds more fine-grained control of the amount of work via \emph{mutation distance}. We can define a mutation distance to be a distance from a test function to a function with the actual mutation. If function A calls function B and function B calls function C, then the distance between A and C is 2. Mutation distance can be used to decrease the number of mutations: Mull can ignore mutations that are too far from a test function.

The instrumentation-based approach has an overhead, but it is necessary to get the right code coverage information. Initially, we used static code analysis to build the call tree: Mull iterated through bitcode and followed call instructions to build the tree. Unfortunately, a function is not always known until runtime. Typical examples are C function pointers and C++ virtual function calls. After a few failed attempts we switched to the dynamic instrumentation. Thus the \emph{call tree} became \emph{dynamic call tree}.

\subsection{JIT and Runtime Compilation}

To run a tested program, Mull needs to compile the bitcode files into object files containing machine code and link them together into executable, as any compiler would do. To accomplish this task Mull utilizes LLVM JIT engine. This approach has a great advantage: compilation and linking happen in memory. Thus there is no disk I/O overhead.

When it comes to mutation Mull performs it on a copy of a single bitcode file, recompiles it and links together with already compiled object files. Partial recompilation helps to increase performance. It also helps to decrease memory usage: mutated bitcode file can be disposed from memory right after execution.

\subsection{Sandboxing}

Mutations can make the code behave in unexpected ways: to crash, to timeout or
to exit prematurely. We use a parent/child process isolation to achieve a
proper sandboxing of a tested program.

Mull, which is a parent process, runs each test in a separate child process.
The \verb;fork; system call is used to create a child process, \verb;mmap; system call is
used to share memory between the parent process and the child process.

Mull handles exit status of a child process according to the following policy:

\subsubsection{Normal execution (test has passed or failed)} \label{sandboxing_normal_execution}
We use a conventional exit
code 227 to indicate if a test has run without any issues. If child process
exits with code 227, Mull knows that a test has either succeeded or failed and
that nothing extraordinary like in one of the following cases has happened.

\subsubsection{Timeout} Mutated code might never finish its execution in a child
process. To handle this case Mull sets an alarm in a child process that exits
with a conventional timeout code 239 after a certain time interval. We use
\verb;ualarm; function to set the alarm.

\subsubsection{Crash} Mutated code can crash with a child process executing it. We use \verb;WIFSIGNALED();
to detect a crash of a child process.

\subsubsection{Abnormal exit} Mutated code exits prematurely from a child process and
this does not let a test to finish. This scenario is a reason
for the existence of the custom exit code 227 from the case 1) because Mull needs to
distinguish between normal exit and abnormal exit from a child process.

\subsection{Dry Run} \label{dry_run}

It is not known in advance how many mutations a project has and how much time does it take Mull to run it. To remove uncertainty, we introduce \emph{dry run} mode. In this mode, Mull collects information about mutations but does not run tests against them. Therefore no partial recompilation and no sandboxing are needed.

Additionally, Mull gives a pessimistic approximation of the run time: it calculates how much time would be needed if each mutant times out. Real execution time is lower than the approximation, but it gives a good hint of expected run time.

\subsection{Test Framework: plugin architecture}\label{test_framework}

Mull can work with any test framework. The only requirement is that Mull can run a single test independently from the other tests in a test suite.

Each test framework plugin consists of two components: \emph{test finder} and \emph{test runner}. Mull uses test finder to find the tests in a bitcode of a tested program, and it uses test runner to run one test according to the calling conventions of a given test framework.

Test finder takes all bitcode files as an input and gives back a list of test functions. Examples: for \texttt{SimpleTestFinder} a test is simply a C function whose name starts with \verb;test;, for \texttt{GoogleTestFinder} a search algorithm extracts the information about the test functions from \verb;internal::MakeAndRegisterTestInfo; registration call of a GoogleTest framework.

Test runner runs one test and returns the result of its execution. Running a test can be as easy as calling a test function and checking its return value: for \texttt{SimpleTestRunner} test passes if its test function returns 1 and fails
if it returns 0. For GoogleTest framework \texttt{GoogleTestRunner} has to emulate the work of GoogleTest's \verb;main(); function: to run one test \texttt{GoogleTestRunner} runs a test suite in a "focused mode" with a filter set to a name of this test's function (\verb;--gtest_filter=TestFunctionName;).

Many projects have their custom test suites. Examples are Musl, OpenSSL, glibc, openlibm. While it is possible to create a dedicated pair of test finder and test runner for each of these projects like \texttt{OpenSSLTestFinder} and \texttt{OpenSSLTestRunner}, we created a general solution called \texttt{CustomTestFramework} to enable testing of these projects. To use \texttt{CustomTestFramework} with a given project one has to provide the custom test definitions in a configuration file. Here is an example for OpenSSL project:

\begin{verbatim}
test_framework: CustomTest
custom_tests:
  - name: test_bio_enc_aes_128_cbc
    method: test_bio_enc_aes_128_cbc
    program: bio_enc_test
    arguments: [ test_bio_enc_aes_128_cbc ]
\end{verbatim}

In this case \texttt{CustomTestFinder} treats a function called \verb;test_bio_enc_aes_128_cbc; as a test, and \texttt{CustomTestRunner} runs the program using specified arguments.

\subsection{Fail Fast mode} \label{fail_fast_mode}

In the worst case a tested program with \verb;N; tests and \verb;M; mutations requires \verb;N * M; test runs. Mull has an option to decrease the amount of test runs: \emph{fail fast} mode. For example, if a mutation is reached from 20 tests and the very first test kills the mutant, then there is no need to run the remaining 19 tests. The fail fast mode is disabled by default and can be enabled in the configuration file.

\subsection{Caching} \label{cahing}

Mull uses JIT and Runtime Compilation for better performance. However, sometimes it is faster to read object file from disk than to compile it in memory from LLVM Bitcode. In this regard, Mull supports on-disk cache. Before compiling a bitcode file Mull attempts to get an object file from disk. If there is none, then Mull compiles the bitcode file and saves resulting object file on-disk for later usage. When Mull runs next time, it can use an object file from the previous session.

To avoid a use of outdated object files, Mull encodes checksum of original bitcode file into the name of a cached object file. Object files for mutants also contain a unique identifier of the mutation point.

\section{Supported mutation operators}

Mull performs mutations on the LLVM IR code, so its implementation of mutation operators is largely determined by the specification of LLVM language and in particular its Instruction Reference \cite{LLVMLangReference}. We also used Pitest's documentation of its mutation operators \cite{PitestDocMutationOperators} to decide which operators to implement first.

The following is the list of mutation operators currently supported by Mull. All of the instructions referenced below can be found in the LLVM IR language manual.

\subsection{Math: Add, Sub, Mul, Div}
\label{math_add}

This group of operators performs mutations of basic arithmetic operators: "+" to "-",  "-" to "+",  "*" to "/", "/" to "*".

Math Add replaces an \verb;add; instruction, which returns the sum of its two operands, with a \verb;sub; instruction, which returns the difference of its two operands. Math Add also works with the floating equivalent of \verb;add; instruction: \verb;fadd; which is replaced with \verb;fsub;.

Math Sub operator performs the same kind of mutation as Math Add but in opposite direction: from \verb;sub; to \verb;add; and \verb;fsub; to \verb;fadd;. Math Mul and Div work with \verb;mul;, \verb;fmul; and \verb;div;, \verb;fdiv; instructions respectively.

\subsection{Negate Condition}
\label{negate_condition}

Negate Condition operator works with \verb;icmp; instruction (comparison of integer operands) and \verb;fcmp; instruction (comparison of floating-point operands). Both instructions accept three operands of which \textit{"the first operand is the condition code indicating the kind of comparison to perform"}. This first operand is a conventional code that represents a type of comparison, for example: "unsigned equal" to "signed less than". Negate Condition modifies the code to achieve a complete negation of a condition: from "equal" to "not equal",  from "signed less" to "signed greater than or equal", etc.

\subsection{Remove Void Function}
\label{remove_void_function}

This operator removes a call to a void function from LLVM IR code. The void function calls can be represented by two instructions in LLVM IR: \verb;call; and \verb;invoke;. The difference between these instructions is related to the details of exception handling and is hidden well behind LLVM IR API making this difference irrelevant to Mull.

\subsection{Replace Call}

This operator replaces a function call, whose return value is an integer or floating-point scalar value, with an arbitrary value according to the following simple rule: the function call is replaced with a value forty-two (42) of a corresponding integer or floating-point type. Like Remove Void Function operator, this operator works with \verb;call; and \verb;invoke; instructions.

\subsection{Scalar Value Replacement}

This operator replaces an integer or floating-point scalar value with a predetermined value according to the following simple rule: non-zero value is replaced with a zero value (0) of the corresponding integer or floating-point type, zero value is replaced with a value of one (1) of the corresponding integer or floating-point type. Scalar values can appear as operands of many different instructions in LLVM IR language: binary arithmetic instructions like \verb;add; or \verb;mul;, comparison instructions \verb;icmp; and \verb;fcmp;, return instruction \verb;ret;, function call instructions \verb;call; and \verb;invoke; and some others. Scalar Value operator maintains a list of such instructions that determines if a particular instruction can be a target of a Scalar Value mutation.

\section{Evaluation} \label{evaluation}

In this section, we describe our experience of applying Mull on real-world projects. We focus on ease of integration, performance, and a practical applicability of Mull, rather than on concrete results such as found bugs or shallow tests. For this paper we picked three open-source projects: RODOS \cite{RODOS}, OpenSSL \cite{OpenSSL} and LLVM \cite{LLVM}. Table \ref{projects_table} describes some properties of these projects. The number of lines of code represents size and scale of a project. However, more representative metric is a number and an overall weight of bitcode files: it has a direct impact on performance because all this code has to be loaded into memory, analyzed, compiled, and linked together.

Measurements for OpenSSL and LLVM were made on macOS 10.13 with 16GB of RAM and Intel i7 3.1GHz CPU. Measurements for RODOS were made on the same machine, but inside of VirtualBox running Ubuntu 16.04, 32 bit. 4GB of RAM and two cores of the Intel i7, 3.1GHz were allocated for the virtual machine.

For this experiment we used three mutation operators: Math Add (\ref{math_add}), Negate Condition (\ref{negate_condition}), and Remove Void Function (\ref{remove_void_function}). All tests were run with the Fail Fast mode (\ref{fail_fast_mode}) and Caching (\ref{cahing}) enabled. We ran each group of tests twice: a cold run, without cache in place, and a hot run, with cache in place.

For each project we measure how many tests a test suite has, how many mutants Mull detects given the mutation operators mentioned above, total amount of test runs executed during analysis, and the total execution time for both cold and hot runs.

\begin{table}[!t]
\caption{Projects}
\centering

\begin{tabular}{|c|c|c|c|c|}
\hline
        & Lines     & Bitcode & Bitcode & Average time \\
Project & of code   & files   & size    & per test run \\
\hline                                             
RODOS   & 125,127   & 32      & 407 KB  & 23 ms        \\
\hline                                             
OpenSSL & 311,293   & 630     & 11 MB   & 42 ms        \\
\hline
LLVM    & 1,324,567 & 224     & 242 MB  & 31 ms        \\
\hline
\end{tabular}
\label{projects_table}
\end{table}

\subsection{RODOS}

RODOS \cite{RODOS} is a real-time operating system developed by the German Aerospace Center. It is written in C and C++ and uses CppUnit test framework \cite{CPPUnit} for its test suite. Among several bare-metal platforms, it can be run on Linux and other POSIX-compliant operating systems.

RODOS has many small test suites, each of them covering very specific part of the system. Examples are: \texttt{matrix4d\_test}, \texttt{quaternion\_test}, \texttt{filesystem\_test}, \texttt{hal\_gpio\_test}. Each test suite is designed to run only one single test per compilation: to run a test one has to compile the test suite enabling a test by providing a macro definition. We have to change this to control the test selection at runtime rather than at compile time. Once the test suite is compiled, it can run either all tests, or the one specified via command-line arguments (\verb;argv;). Original test driver always exits with exit code 0. To check if tests failed or not one has to either observe the output or check a test report that is written into XML file. We have to add a small change here as well: exit code should represent amount of failed tests. If all tests pass, then exit code is 0, otherwise some positive number. This is a widely used approach. RODOS uses Makefiles to build and run its test suites, we have to change compiler flags (\verb;CXXFLAGS;) to emit LLVM Bitcode.

Two more workarounds are required to run Mull against RODOS. Some parts of the system are written in assembly so they are compiled directly into machine code. We have to point Mull to them using \verb;object_file_list; configuration option. Since RODOS uses CppUnit we also have to point it to the \verb;libcppunit.so; via \verb;dynamic_library_list; configuration option.

Once these preparations are done, we can run Mull against RODOS. For this experiment, we pick five test suites based on amount of tests in each of them. Table \ref{rodos_table} contains the results.

\begin{table}[!t]
\caption{Results for RODOS}
\centering

\begin{tabular}{|c|c|c|c|c|c|}
\hline
                   &        &          &  Test   &           & Total time     \\
Test suite         & Tests  &  Mutants &  runs   &  Distance & (cold / hot)   \\
\hline

linkinterfaceuart  & 11     & 19       & 186     & 2         & 4s / 2s        \\
\hline
stdlib\_pico       & 10     & 36       & 356     & 2         & 5s / 4s        \\
\hline
thread       		& 10     & 57       & 196     & 3         & 5s / 3s        \\
\hline
sortedlist       	& 9      & 16       & 85      & 4         & 2s / 2s        \\
\hline
linkinterfacecan   & 8      & 18       & 471     & 5         & 19s / 12s      \\
\hline
\end{tabular}
\label{rodos_table}
\end{table}

\subsection{OpenSSL}

OpenSSL \cite{OpenSSL} is a well-known implementation of TLS and SSL protocols. It is written in C. It uses custom test framework for its tests. OpenSSL has a mix of unit and integration tests. We have to look at each test suite to identify if it is a unit test suite or an integration test suite. Test suites with unit tests are simple programs that are compiled into an executable. Each test suite can only run all tests at once. We have to change them to run one test based on command line arguments, or to run all of them if no arguments are given, to preserve original behavior. We also have to extract information about each test manually. To set up \texttt{CustomTestFramework} Mull needs to know which function is a test and which command line arguments to pass to run this very specific test.

Obtaining LLVM Bitcode is trivial. To compile OpenSSL one has to invoke \verb;configure; script to prepare build system. \verb;configure; accepts additional parameters that are used as \verb;CFLAGS;. We invoke the script by passing \verb;-flto; to enable LTO \cite{LLVM_LTO}, which produces bitcode files instead of object files as build artifacts. Then we compile test suites of interest and construct separate configuration files for each of them.

Results of this experiment can be found in Table \ref{openssl_table}.

\begin{table}[!t]
\caption{Results for OpenSSL}
\centering

\begin{tabular}{|c|c|c|c|c|c|}
\hline
               &         &          &  Test   &           & Total time     \\
Test suite     & Tests   &  Mutants &  runs   &  Distance & (cold / hot)   \\
\hline

packettest     & 22      & 67       & 173     & 3         & 30s / 14s      \\
\hline
destest        & 20      & 274      & 1256    & 3         & 47s / 29s      \\
\hline
test\_test     & 19      & 102      & 214     & 4         & 31s / 17s      \\
\hline
igetest        & 10      & 118      & 709     & 2         & 29s / 18s      \\
\hline
bio\_enc\_test & 6       & 708      & 2667    & 12        & 3m8s / 2m45s   \\
\hline

\end{tabular}
\label{openssl_table}
\end{table}

\subsection{LLVM} \label{evaluation_llvm}

The LLVM \cite{LLVM} compiler infrastructure project is the biggest project we have analyzed so far. LLVM is written in C++. It uses GoogleTest \cite{GoogleTest} as a test framework. It has several unit test suites of various sizes targeting different subsystems of LLVM. For this experiment, we use ADTTests: a test suite that covers specific abstract data types used in LLVM such as arrays, strings, maps, integers, floats, etc. Additionally, in this test suite, we focus only on normal tests and exclude tests which are based on Typed Tests feature of GoogleTest that \verb;GoogleTestFinder; does not support yet.

Obtaining Bitcode is trivial for LLVM: it has a build setting that enables LTO \cite{LLVM_LTO}, which produces bitcode files instead of object files as build artifacts. We use only one workaround to get LLVM's tests running: LLVM JIT does not support Thread-Local Storage \cite{LLVMBugTLS} so we have to exclude one source file that uses TLS from the compilation. Fortunately, this file is not used in the test suite so its absence does not affect the analysis.

LLVM is a big project. It this case it is recommended to launch Mull in Dry Run mode (\ref{dry_run}) to get information about tested program. Dry run shows that Mull detected 550 tests and found 11779 mutants, it also shows that there are 60325 test runs according to \ref{fail_fast_mode}. The report also shows approximation of execution time: full run may take about 9 hours at maximum. It helps to see the order: hours, not days in the worst case. The approximation is very pessimistic: Mull assumes that every mutant fails because of a timeout. In fact, real execution time was 3 hours 46 minutes.

Table \ref{adttests_table} shows the results for different configurations. We use three groups of tests of different size and different mutation distance (\ref{dynamic_call_tree}) for each of them to show applicability of Mull even for big projects. The execution time of almost four hours on the whole test suite is impractical for iterative development process as opposed to two minutes for a subset of tests.

\begin{table}[!t]
\caption{Results for LLVM}
\centering

\begin{tabular}{|c|c|c|c|c|c|}
\hline
              &         &          &  Test   &           & Total time     \\
Test suite    & Tests   &  Mutants &  runs   &  Dist.    & (cold / hot)   \\
\hline

All Tests     & 550     & 11779    & 60325   & 25        & 3h46m / 1h54m  \\
\hline
All Tests     & 550     & 5508     & 13601   & 2         & 1h52m / 47m46s \\
\hline

APFloat       & 70      & 1894     & 22010   & 25        & 41m1s / 18m21s \\
\hline
APFloat       & 70      & 361      & 1622    & 2         & 14m13s / 4m32s \\
\hline

StringExtras  & 5       & 160      & 165     & 7         & 4m44s / 3m2s   \\
\hline
StringExtras  & 5       & 93       & 98      & 2         & 4m36s / 2m24s  \\
\hline

\end{tabular}
\label{adttests_table}
\end{table}

\section{Current limitations}

\subsection{Junk and stray mutations}

A mutation can exist in bitcode, but cannot be achieved by changing original source code. Such mutation is called \emph{junk mutation}. The term was first coined by Henry Coles \cite{junk_mutation_reference}. A good example of such mutation in C++ is a \verb;std::vector::push_back; method call: one line of C++ code produces around 200 LLVM IR instructions. Depending on mutation operator Mull finds mutations in those instructions even though there is no equivalent in the original source code. Some mutation operators require advanced pattern matching to avoid this issue, for others, we did not find a robust solution yet.

Another issue is C and C++ code from their standard libraries. Compiler inlines code from macros and templates into resulting bitcode. Mull finds mutations in this code as well, but they are not relevant to a tested program. We call such mutations \emph{stray mutations}. Fortunately, there is a simple workaround: Mull can filter out mutations based on their location in a source code using \verb;exclude_locations; configuration option. This approach also helps to avoid mutation of third-party code.

\subsection{Current limitations of LLVM JIT}

Mull uses LLVM JIT from which it gets its power as well as some of its limitations. The following are two major limitations we encountered: LLVM JIT does not work with projects using Thread Local Storage \cite{LLVMBugTLS}, and it does not support Objective-C Runtime \cite{LLVMJITObjC}. The latter limitation is the only reason why Mull does not yet fully support Objective-C and Swift programming languages. Both problems are solvable and are waiting for their solution.

\section{Future work}

There is a lot of work to be done to get Mull closer to its use in production. Below, we outline the three major (and most obvious) parts of our work: performance improvements, integration with modern IDE's, further exploration of the real-world projects. 

One direction of work is further performance optimizations: parallelization and even better control over recompilation of bitcode. Mull still runs only one child process at a time so the work with multiple child processes is one of the nearest optimizations we are planning. Recompilation of mutated function instead of a whole bitcode file that contains it can improve performance of Mull on projects with large bitcode files.

Integration with existing IDE's is yet another important part of work to make Mull practical for daily use. While Mull works perfectly as a command-line tool that produces HTML reports, we also see it natural to be a part of a workflow provided by the modern IDE's.

Another direction of work is a further exploration of the real-world projects that will drive the implementation of new test framework adapters like Catch for C++, better support of programming languages like Rust and Swift, running Mull on BSD and Windows systems. In this regard, we especially look forward to the proper support of Objective-C Runtime by LLVM JIT because it will open Mull the door to the world of desktop and mobile application development on macOS and iOS platforms.

We are aware that other mutation testing tools for compiled programming languages exist \cite{MuCPP} and we assume that a proper comparison between Mull and these tools should be a topic of separate research.
\section{Conclusion}

Our choice of LLVM as a base for an implementation of a mutation testing tool was based on an experiment with LLVM IR and LLVM JIT libraries that had produced results superior to those from any of our previous attempts to implement a solution working on source code or AST levels. So far, we did not encounter a single critical problem that would turn us away from our decision to base Mull on LLVM with its intermediate language and infrastructure. Quite to the opposite, Mull satisfies all criteria that we consider important for an implementation of mutation testing tool. It has a great number of applications and a large room for further improvement.

To test Mull on real-world projects and to explore possible limitations of our approach we applied it to as many different projects, programming languages, test frameworks and operating systems as was possible with our capacity. The following is the list of the projects we analyzed:
\begin{itemize}
\item \textit{LLVM} (C/C++, GoogleTest, macOS)
\item \textit{OpenSSL} (C/C++, custom test suite, macOS)
\item \textit{RODOS} (C/C++, CppUnit, Linux)
\item  \textit{openlibm} (C, custom test suite, macOS)
\item  \textit{newlib's libm} (C, custom test suite, Linux)
\item  \textit{fmt} (C++, GoogleTest, macOS)
\item  \textit{CryptoSwift} (Swift, XCTest, Linux)
\item  \textit{rustc-demangle} (Rust, Rust's test framework, macOS)
\item  \textit{Mull (autoanalysis)} (C++, GoogleTest, macOS)
\end{itemize}

Our long-term goal is to get Mull to the point where it can be used by industry as a drop-in solution for mutation testing. Also, we expect Mull to find a use in research, including interaction with other tools and approaches, that would find solutions to speed up the normally slow process of mutation testing with automatic test generation.

\section*{Acknowledgements}

We thank Henry Coles and Markus Schirp for fruitful discussions and their helpful advice at the early stage of development of Mull.

We thank Yue Jia and Mark Harman for their Analysis and Survey \cite{MutationTestingSurvey} that gave us a theoretical background for our work.

We thank Tobias Grosser for the hint about LTO option that helps to get LLVM Bitcode from a project's source code.

\bibliography{references}{}

\begin{thebibliography}{10}
\providecommand{\url}[1]{#1}
\csname url@samestyle\endcsname
\providecommand{\newblock}{\relax}
\providecommand{\bibinfo}[2]{#2}
\providecommand{\BIBentrySTDinterwordspacing}{\spaceskip=0pt\relax}
\providecommand{\BIBentryALTinterwordstretchfactor}{4}
\providecommand{\BIBentryALTinterwordspacing}{\spaceskip=\fontdimen2\font plus
\BIBentryALTinterwordstretchfactor\fontdimen3\font minus
  \fontdimen4\font\relax}
\providecommand{\BIBforeignlanguage}[2]{{%
\expandafter\ifx\csname l@#1\endcsname\relax
\typeout{** WARNING: IEEEtran.bst: No hyphenation pattern has been}%
\typeout{** loaded for the language `#1'. Using the pattern for}%
\typeout{** the default language instead.}%
\else
\language=\csname l@#1\endcsname
\fi
#2}}
\providecommand{\BIBdecl}{\relax}
\BIBdecl

\bibitem{GithubTopicsMutationTesting}
\BIBentryALTinterwordspacing
``Github topics: Mutation testing.'' [Online]. Available:
  \url{https://github.com/topics/mutation-testing}
\BIBentrySTDinterwordspacing

\bibitem{PITEST}
\BIBentryALTinterwordspacing
H.~Coles, ``Pitest.'' [Online]. Available: \url{http://pitest.org}
\BIBentrySTDinterwordspacing

\bibitem{MUTANT}
\BIBentryALTinterwordspacing
M.~Schirp, ``Mutant.'' [Online]. Available: \url{https://github.com/mbj/mutant}
\BIBentrySTDinterwordspacing

\bibitem{LLVMPaper}
\BIBentryALTinterwordspacing
C.~Lattner and V.~Adve, ``Llvm: A compilation framework for lifelong program
  analysis \& transformation,'' in \emph{Proceedings of the International
  Symposium on Code Generation and Optimization: Feedback-directed and Runtime
  Optimization}, ser. CGO '04.\hskip 1em plus 0.5em minus 0.4em\relax
  Washington, DC, USA: IEEE Computer Society, 2004, pp. 75--. [Online].
  Available: \url{http://dl.acm.org/citation.cfm?id=977395.977673}
\BIBentrySTDinterwordspacing

\bibitem{MullSourceCode}
\BIBentryALTinterwordspacing
A.~Denisov and S.~Pankevich, ``Mull.'' [Online]. Available:
  \url{https://github.com/mull-project/mull}
\BIBentrySTDinterwordspacing

\bibitem{apache2}
\BIBentryALTinterwordspacing
``Apache license,'' Apache Software Foundation. [Online]. Available:
  \url{https://www.apache.org/licenses/LICENSE-2.0}
\BIBentrySTDinterwordspacing

\bibitem{LLVMLangReference}
\BIBentryALTinterwordspacing
``{LLVM} {L}anguage {R}eference {M}anual: {I}nstruction {R}eference.''
  [Online]. Available:
  \url{https://releases.llvm.org/3.9.0/docs/LangRef.html#instruction-reference}
\BIBentrySTDinterwordspacing

\bibitem{PitestDocMutationOperators}
\BIBentryALTinterwordspacing
H.~Coles, ``Pitest: {A}vailable mutation operations.'' [Online]. Available:
  \url{http://pitest.org/quickstart/mutators/}
\BIBentrySTDinterwordspacing

\bibitem{RODOS}
\BIBentryALTinterwordspacing
``{RODOS}.'' [Online]. Available:
  \url{https://en.wikipedia.org/wiki/Rodos_(operating_system)}
\BIBentrySTDinterwordspacing

\bibitem{OpenSSL}
\BIBentryALTinterwordspacing
``{O}pen{SSL}.'' [Online]. Available: \url{https://www.openssl.org}
\BIBentrySTDinterwordspacing

\bibitem{LLVM}
\BIBentryALTinterwordspacing
``{LLVM}.'' [Online]. Available: \url{https://llvm.org}
\BIBentrySTDinterwordspacing

\bibitem{CPPUnit}
\BIBentryALTinterwordspacing
``{C}pp{U}nit.'' [Online]. Available:
  \url{https://sourceforge.net/projects/cppunit/}
\BIBentrySTDinterwordspacing

\bibitem{LLVM_LTO}
\BIBentryALTinterwordspacing
``{LLVM} {L}ink {T}ime {O}ptimization: {D}esign and {I}mplementation.''
  [Online]. Available: \url{https://llvm.org/docs/LinkTimeOptimization.html}
\BIBentrySTDinterwordspacing

\bibitem{GoogleTest}
\BIBentryALTinterwordspacing
``{G}oogle{T}est.'' [Online]. Available:
  \url{https://github.com/google/googletest}
\BIBentrySTDinterwordspacing

\bibitem{LLVMBugTLS}
\BIBentryALTinterwordspacing
``{MCJIT TLS} support: {C}annot select: {X86ISD}::{W}rapper{RIP}.'' [Online].
  Available: \url{https://bugs.llvm.org/show_bug.cgi?id=21431}
\BIBentrySTDinterwordspacing

\bibitem{junk_mutation_reference}
\BIBentryALTinterwordspacing
H.~Coles, ``{J}unk {M}utations.'' [Online]. Available:
  \url{https://twitter.com/0hjc/status/478896988784963584}
\BIBentrySTDinterwordspacing

\bibitem{LLVMJITObjC}
\BIBentryALTinterwordspacing
``[llvm-dev] {I}s it possible to execute {O}bjective-{C} code via {LLVM JIT}?''
  [Online]. Available:
  \url{http://lists.llvm.org/pipermail/llvm-dev/2016-October/106218.html}
\BIBentrySTDinterwordspacing

\bibitem{MuCPP}
P.~Delgado-Pérez, I.~Medina-Bulo, F.~Palomo-Lozano, A.~García-Domínguez, and
  J.~Domínguez-Jiménez, ``Assessment of class mutation operators for c++ with
  the mucpp mutation system,'' vol.~81, p. 169–184, 01 2017.

\bibitem{MutationTestingSurvey}
Y.~Jia and M.~Harman, ``An analysis and survey of the development of mutation
  testing,'' \emph{IEEE Transactions on Software Engineering}, vol.~37, no.~5,
  pp. 649--678, Sept 2011.

\end{thebibliography}
\bibliographystyle{IEEEtran}

\end{document}